\documentclass[]{aa}

\usepackage{graphicx}
\usepackage{color}                       
\usepackage{natbib}
\usepackage{ulem}
\usepackage{setspace}
\usepackage{subfig}

\newcommand{\lapprox} {\, \lower3pt\hbox{$\sim$}\llap{\raise2pt\hbox{$<$}}\,}
\newcommand{\gapprox} {\, \lower3pt\hbox{$\sim$}\llap{\raise2pt\hbox{$>$}}\,}

\begin{document}
\begin{spacing}{1}

\title{Determination of the Acceleration Region Size in a Loop-structured Solar Flare}

   \author{Jingnan Guo\thanks{guo@dima.unige.it}
          \inst{1}
          \and
          A. Gordon Emslie
          \inst{2}
          \and
          Eduard P. Kontar
          \inst{3}
          \and
          Federico Benvenuto
          \inst{1}
          \and
          Anna Maria Massone
          \inst{4}
          \and
          Michele Piana
          \inst{1, 4}
          }

   \institute{	(1) Dipartimento di Matematica, Universit\`a di Genova, via Dodecaneso 35, 16146 Genova, Italy \\
			  	(2) Department of Physics and Astronomy, Western Kentucky University, Bowling Green, KY 42101, USA \\
 				(3) School of Physics and Astronomy, SUPA, University of Glasgow, Glasgow, G12 8QQ, Scotland \\
 				(4) CNR - SPIN, via Dodecaneso 33, I-16146 Genova, Italy
			}
   \date{}

\abstract
{}
{In order to study the acceleration and propagation of bremsstrahlung-producing electrons in solar flares, we analyze the evolution of the flare loop size with respect to energy at a variety of times. A GOES M3.7 loop-structured flare starting around 23:55 on 2002~April~14 is studied in detail using \textit{Ramaty High Energy Solar Spectroscopic Imager} (\textit{RHESSI}) observations.}
{We construct photon and mean-electron-flux maps in 2-keV energy bins by processing observationally-deduced photon and electron visibilities, respectively, through several image-processing methods: a visibility-based forward-fit (FWD) algorithm, a maximum entropy (MEM) procedure and the uv-smooth (UVS) approach. We estimate the sizes of elongated flares (i.e., the length and width of flaring loops) by calculating the second normalized moments of the intensity in any given map. Employing a collisional model with an extended acceleration region, we fit the loop lengths as a function of energy in both the photon and electron domains.}
{The resulting fitting parameters allow us to estimate the extent of the acceleration region which is between $\sim 13~\rm{arcsec}$ and $\sim 19~\rm{arcsec}$. Both forward-fit and uv-smooth algorithms provide substantially similar results with a systematically better fit in the electron domain.}
{The consistency of the estimates from these methods provides strong support that the model can reliably determine geometric parameters of the acceleration region. The acceleration region is estimated to be a substantial fraction ($\sim 1/2$) of the loop extent, indicating that this dense flaring loop incorporates both acceleration and transport of electrons, with concurrent thick-target bremsstrahlung emission.}

\keywords{Sun: flares --- Acceleration of particles }

\maketitle
\authorrunning{Jingnan Guo et al.}
\titlerunning{Sizes of the Acceleration Region}

\section{Introduction}\label{intro}

Solar flares are known to produce large quantities of accelerated particles, in particular electrons in the deka-keV to deci-MeV range.  However, the location and physical properties of the acceleration region are yet to be well constrained.  An intrinsic complication is that the radiation produced by energetic particles emanates not only from the acceleration region itself, but also from other locations in the flare into which the accelerated particles propagate.  Indeed, the oft-used ``thick-target'' model \citep{brown71} exploits this very complication by deriving properties of the hard X-ray emission that are completely independent of the location, extent, or physical properties of the acceleration region. Hence, determination of the properties of the acceleration region from spatially-integrated observations of flare emission is not straightforwardly possible. The reader is referred to recent reviews on electron properties inferred from hard X-rays \citep{kontar2011review} and their implications for electron transport \citep{holman2011review}.

With the availability of high-quality hard X-ray imaging spectroscopy data from the \textit{RHESSI} instrument \citep{linetal02}, the situation has much improved \citep[see, e.g.,][]{emslie2003rhx}. Higher energy electrons are able to propagate further from the acceleration region and hence produce hard X-ray emission over a greater spatial extent than in lower-energy bands. \citet{xuetal08} and \citet{kohabi11} analyzed a set of events characterized by simple coronal flare loop sources located near the solar limb. In order to determine the spatial properties of the flare loops they fitted the {\textit{RHESSI}} visibilities with the geometric parameters of the loops and determined the size of the acceleration regions by fitting the source extents as a function of the photon energy with a collisional acceleration and propagation model.

The present paper extends this kind of analysis. Specifically, for the simple coronal loop event observed by \textit{RHESSI} on 2002~April~14, we study the variation of source extent with energy not only in the {\textit{photon}} energy domain, but also, for the first time, in the {\textit{electron}} domain, using the procedure for generating weighted mean electron flux maps first enunciated by \cite{pianaetal07}. This extension of the model to the electron domain not only admits a simpler description of the source size with energy $E$ but also allows us to exploit the ``rectangular'' nature of the spectral inversion process (in bremsstrahlung, photons with a given energy $\epsilon$ are produced by electrons with all higher energies $E \geq \epsilon $), making it possible to reconstruct electron maps at energies higher than the maximum photon energy observed \citep[see][]{kontar2004}. Moreover, the regularization algorithm used to invert count visibilities into electron visibilities introduces a natural smoothing over energy $E$, which leads to a smoother behavior of source size with energy and so a more reliable estimate of pertinent parameters such as the acceleration region length.

For a given time interval, we use both photon and electron maps to examine the form of the variation of loop length with energy. The analysis employs three different visibility-based imaging algorithms for both photon and electron maps: visibility forward-fit \citep{schmahl2007}, maximum entropy \citep{cornwell1985, bong2006}, and uv-smooth \citep{massone2009}. In the visibility-forward-fit procedure, the loop sizes are determined as model parameters. For the other methods, we use the standard deviation -- square root of the second normalized moment -- of the intensity map as a measure of the loop half-length. We then use the "tenuous acceleration region" model of \cite{xuetal08} to derive the longitudinal and lateral extents of the acceleration site.  In Section~\ref{visibilities}, we describe the inversion algorithms used to derive electron flux visibilities and the imaging techniques employed to create the corresponding electron maps for a given time interval and electron energy range. In Section~\ref{data}, we fit the variation of loop size with electron energy $E$ to a simple parametric model \citep{xuetal08} in order to determine the longitudinal and lateral extents of the acceleration region. In Section~\ref{result}, we compare the values obtained through different imaging techniques and from different map domains (photon and electron).

The inferred length of the acceleration region ($\sim 15$~arcsec) is approximately half the total length of the loop.  This suggests that the standard model of solar flares where electrons are initially accelerated at a reconnection site near/above the loop top \citep[e.g.,][]{kopp1976} is not appropriate for certain types of flares.  In such flares, the acceleration instead takes place over a large region inside the flare loop.

\section{\textit{RHESSI} Visibilities and Imaging Processes}\label{visibilities}

Solar flare hard X-ray emission is principally produced by accelerated electrons through the bremsstrahlung process \citep{brown71}. The relation between the mean electron flux spectrum ${\overline F}(x,y;E)$ (electrons cm$^{-2}$~s$^{-1}$~keV$^{-1}$), averaged over the line-of-sight direction $z$ through the point $(x,y)$ in a target source, and the corresponding bremsstrahlung hard X-ray spectrum $I(x,y;\epsilon)$ (photons cm$^{-2}$~s$^{-1}$~keV$^{-1}$~arcsec$^{-2}$) emitted per unit area of the source, can be written as \citep{pianaetal07}

\begin{equation}
I(x,y; \epsilon) = {a^2 \over 4 \pi R^2}\int^{\infty}_{E
=\epsilon} N(x,y) \, {\overline F}(x,y; E) \, Q(\epsilon, E) \,
dE,\label{fund_app}
\end{equation}
where $a = 7.25 \times 10^7$~cm~arcsec$^{-1}$, representing the extent of a one arcsec source at a distance $R = 1$~AU; $Q(\epsilon, E)$ (cm$^2$~keV$^{-1}$) is the bremsstrahlung cross-section for emission of a photon at energy $\epsilon$; and $N(x,y)$ (cm$^{-2}$) is the column density along the line-of-sight direction.

By recording the temporal modulation of the detected flux passing through sets of rotating absorbing grids, \textit{RHESSI} \citep{linetal02} encodes imaging information in terms of a set of spatial Fourier components of the source, termed {\it visibilities}, distributed over nine circles in the spatial frequency $(u,v)$ plane. We define the {\it count visibility spectrum} $V(u,v;q)$ (counts~cm$^{-2}$~s$^{-1}$~keV$^{-1}$) as the two-dimensional spatial Fourier transform of the count spectrum image $J(x,y;q)$ (counts~cm$^{-2}$~s$^{-1}$~keV$^{-1}$~arcsec$^{-2}$).
Similarly, the {\it electron flux visibility spectrum} $W(u,v;E)$ (electrons~cm$^{-2}$~s$^{-1}$~keV$^{-1}$) represents the two-dimensional spatial Fourier transform of the line-of-sight-column-density-weighted mean electron flux image $N(x,y) \, {\overline F}(x,y;E)$ (electrons~cm$^{-4}$~s$^{-1}$~keV$^{-1}$).

The relation between the observed count visibility spectrum $V(u,v;q)$ and the electron flux visibility spectrum $W(u,v;E)$ is \citep{pianaetal07}:

\begin{equation}
V(u,v;q) = {1 \over 4 \pi R^2} \, \int_q^{\infty} W(u,v;E) \, K(q,E) \, dE .
\label{result_red_app}
\end{equation}
Here the kernel $K(q,E)$ satisfies

\begin{equation}
K(q,E) \, dq = \int_{\epsilon = q}^\infty D(q,\epsilon) \,
Q(\epsilon, E) \, d \epsilon \, , \label{ce_cross}
\end{equation}
where $D(q,\epsilon)$ is the detector response matrix.

In order to invert Equation~(\ref{result_red_app}) to obtain the electron flux visibility spectrum $W(u,v;E)$ from the observed count visibility spectrum $V(u,v;q)$, we employed a Tikhonov regularization technique, which has been proven to be a robust and effective inversion method that results in visibilities (and so images) that vary smoothly with electron energy $E$.  Then, from either photon or electron visibility sets, we produce the corresponding images using the visibility-forward-fit, maximum entropy and the uv-smooth interpolation/extrapolation methods, as described in Sections~\ref{fwd} through~\ref{uvs}, respectively.

\subsection{Visibility-based Forward-fit Algorithm}\label{fwd}

The visibility-based forward-fit (FWD) imaging algorithm \citep{schmahl2007} assumes a parametric source form and determines the values of the model parameters that result in the best fit to the visibility data. This method provides not only quantitative values of the parameters but also their uncertainties.  Although the applicability of the FWD algorithm rapidly deteriorates for complex flare morphologies, because of the relatively large number of parameters required to characterize the source structure adequately, the FWD approach is rather effective for sources with a relatively simple structure such as the one we study here.

The FWD routines embedded in the \textit{RHESSI} SolarSoftWare (SSW) provide four simple parametric source geometries: a circular-Gaussian-distributed single source, multiple Gaussian sources, an elliptical-Gaussian source, and a curved-elliptical-Gaussian loop \citep{Hurford2002, schmahl2007}. Since the flare we are analyzing has a simple loop-structured geometry, we adopt the curved elliptical Gaussian form. One of the parameters determined by this routine is the FWHM, which for a Gaussian profile, is related to the standard deviation $\sigma$ by $\exp[-(FWHM/2)^2/2\sigma^2] = 1/2$, i.e., $\sigma = FWHM/\sqrt{8 \ln 2}$.

We applied this method to visibility data in both the photon and electron domain to obtain the standard deviations $\sigma(\epsilon)$ and $\sigma(E)$ in the photon and electron domains, respectively. Since photons of energy $\epsilon$ are produced by electron of all energies $E \ge \epsilon$, these two standard deviations are not the same; indeed, one expects in general $\sigma(\epsilon=E) > \sigma(E)$ (see equations~[\ref{alphadef}] and~[\ref{betadef}]).

\subsection{Maximum Entropy Method}\label{mem}

The basis of the Maximum Entropy Method is to maximize the information entropy while minimizing the $\chi^2$ of fit and maintaining the correct value of the total flux. In the MEM-NJIT method \citep{bong2006}, the visibility amplitudes $V$ are used to calculate the overall flux of the map.  The method implements a statistical regularization method where the functional

\begin{eqnarray}\label{c1}
J = H - \alpha \chi^2\\
 \qquad H = - \sum_j^{n_{pix}} \frac{I_j}{\sum_{j}^{n_{pix}} I_j} \log \left(\frac{I_j}{\sum_{j}^{n_{pix}} I_j} \right) \\
  \qquad \chi^2 = \sum_{l=1}^{n_{vis}} \frac{|V_l -
V_l^{\prime}|}{\sigma_l^2}
\end{eqnarray}
is minimized by means of an iterative scheme.

In these equations, $I$ is the X-ray image made of $n_{pix}$ pixels, $V$ and $V^{\prime}$ are the observed and predicted visibilities, respectively, and the $\sigma$ denote the standard deviations associated with each visibility.  The regularization parameter $\alpha$ is obtained by means of optimization techniques. The MEM algorithm implemented in SSW \citep{bong2006} provides reliable reconstructions in the case of compact events, although the method often super-resolves the sources and can present convergence problems.

\subsection{Fourier-based UV-Smooth Imaging}\label{uvs}

An alternative visibility-based imaging method, termed the {\it uv-smooth} (UVS) algorithm, has been recently developed by \citet{massone2009} and is available on the SSW tree. This method first interpolates the sparsely distributed visibilities to generate a smooth continuum of Fourier components in the spatial-frequency $(u,v)$ plane. Then it performs an FFT-based constrained iterative algorithm to obtain out-of-band extrapolations. The method has proven to reproduce realistic forms of the source structure with a high degree of accuracy, fidelity, robustness, and computational efficiency. Although the method may introduce artifacts when applied to source configurations characterized by distant footpoints, uv-smooth is very accurate when reconstructing relatively localized extended sources such as those considered here.

Unlike images produced by the FWD algorithm, the UVS and MEM procedures do not straightforwardly provide quantitative information on the uncertainties in the determined source extents. Given the electron flux ${\overline F}(x, y;E)$ in a two-dimensional flux image at any given energy $E$, the {\it location} of the source can be estimated by calculating the first normalized moment of the intensity.  The length and width of a source, in the photon and electron domains respectively, can be found by considering the pertinent second normalized moments:

\begin{eqnarray}\label{second-moment}
\sigma(\theta;\epsilon) = \sqrt{\frac{\int_{0}^{\infty} s^2 I(s, \theta;\epsilon) \, ds}{\int_{0}^{\infty} I(s, \theta;\epsilon) \, ds}} \\
 \qquad \sigma(\theta;E) = \sqrt{\frac{\int_{0}^{\infty} s^2 N(s) \, {\overline F}(s, \theta;E) \, ds}{\int_{0}^{\infty} N(s) \, {\overline F}(s, \theta;E) \, ds}} ,
\end{eqnarray}
where $(s,\theta)$ are polar coordinates in the plane of the sky, relative to an origin that we define as the location of the maximum flux intensity. The integrals in Equation~(\ref{second-moment}) are computed numerically for a variety of $\theta$ values.  The maximum (minimum) values of $\sigma(\theta)$ can be identified as the length (width) of the source. In order to provide quantitative uncertainties on these values, we applied a Monte Carlo approach in which random noise is added to the visibilities and the resulting images recomputed and reanalyzed.  Ten realizations of the source visibilities were used; the standard deviations of these ten results were taken to be the 1$\sigma$ errors of the loop length and width.

\section{Application to A Loop-Structured Flare}\label{data}

\begin{center}
\begin{figure}[pht]
\begin{tabular}{cc}
\subfloat[a] { \includegraphics[width=0.45\textwidth]{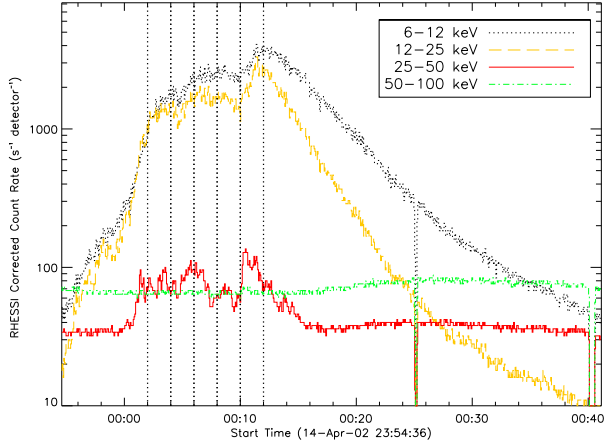} }\\
\subfloat[b] { \includegraphics[width=0.45\textwidth]{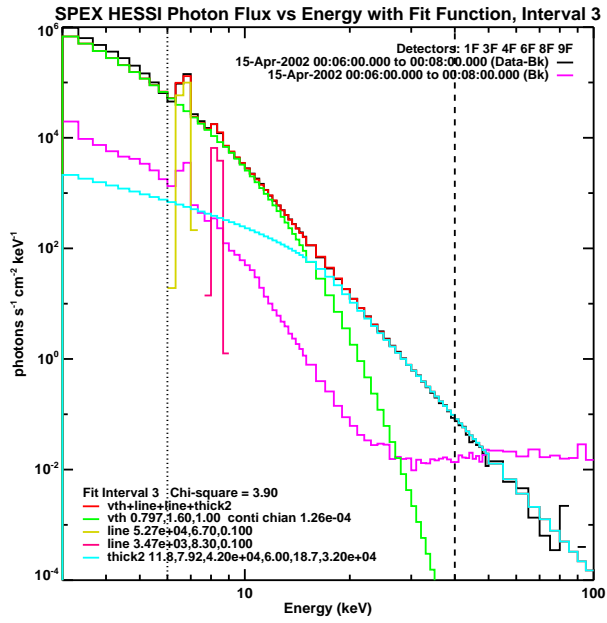} } \\
\subfloat[c] { \includegraphics[width=0.45\textwidth]{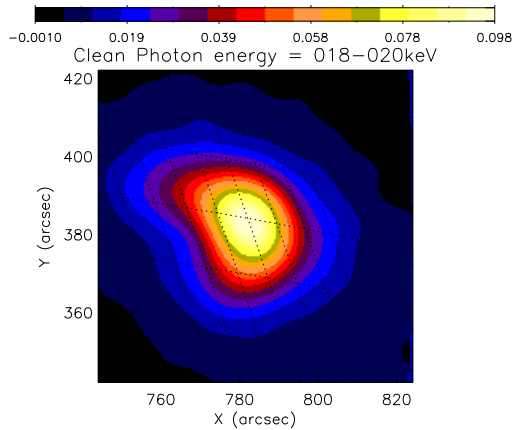} }\\
\end{tabular}
\caption{[a]: Time profile in different \textit{RHESSI} energy ranges of the flare starting around 23:55~UT on 2002~April~14. The vertical lines indicate the five time intervals for which the spectral fitting and visibility-based imaging were performed. [b]: The spatially integrated \textit{RHESSI} X-ray spectrum fitted with an isothermal plus thick-target nonthermal model for the third time interval (00:06 - 00:08 UT). [c]: An example of a photon flux map (for the third time interval and energy bin 18-20~keV) obtained by the Clean method \citep{Hurford2002}.}
\label{fig:ltc_map}
\end{figure}
\end{center}

In the standard flare model, thick-target ``footpoints'' are considered to represent the dominant locations of hard X-ray emission because the coronal magnetic loops through which the energetic particles propagate are generally not dense enough to stop electrons via Coulomb collisions.  Due to the high density of the chromosphere, the hard X-ray structure of footpoint sources typically extends over a very small spatial extent, so that observations with available spatial resolutions cannot directly determine details of the particle acceleration and propagation processes in the bremsstrahlung-emitting region.

However, \textit{RHESSI} has revealed a new class of flares in which the hard X-ray emission is predominantly from the coronal loop itself \citep{krucker08, vebr04, suetal04}.  For such sources, the corona is not only the location of particle acceleration, but also dense enough to act as a thick target, stopping the accelerated electrons before they can penetrate to the chromosphere. Close investigations of these flares can provide direct information on the electron acceleration and propagation processes in the bremsstrahlung-emitting region.

One of the most closely studied events from this class is the ``midnight'' flare of 2002~April~14 \citep{suetal04, vebr04, boetal07, xuetal08, kohabi11}. Figure~\ref{fig:ltc_map}a shows the \textit{RHESSI} count rate profiles for this event in five different energy channels.
For its entire duration this flare consists of a simple loop structure viewed ``side-on'' near the solar limb, as shown in Figure~\ref{fig:ltc_map}c.
The coronal loop has a density sufficiently high \citep[$n \simeq 10^{11}$~cm$^{-3}$;][]{vebr04} to stop electrons up to $\sim 30$~keV over the observed loop length of $10^9$~cm.  It has also been suggested that electrons must be continuously accelerated during this event because for such a dense loop 20~keV particles have collisional lifetimes less than 0.1~s \citep{kohabi11}.

\begin{center}
\begin{figure}[pht]
\begin{tabular}{cc}
 \includegraphics[width=0.50\textwidth]{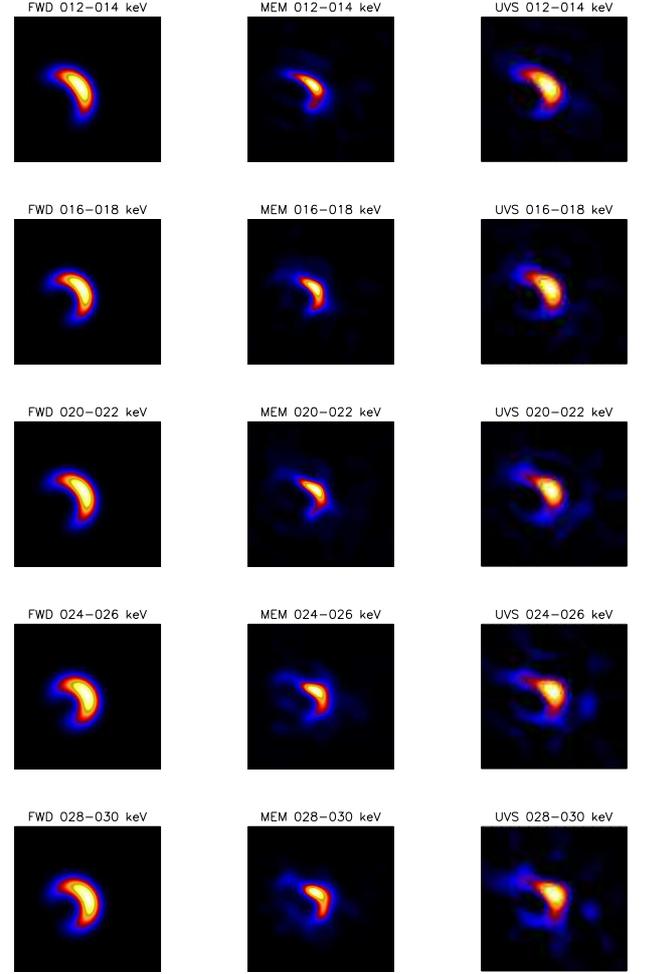}
\end{tabular}
\caption{Mean electron flux maps for the time interval 00:06 - 00:08~UT, obtained through visibility-forward-fit ({\it left }), Maximum Entropy (MEM-NJIT) ({\it middle}) and uv-smooth ({\it right}) procedures applied to electron visibilities. Energy bins are from 12-14~keV ({\it top}) to 28-30~keV ({\it bottom}). The range for x-axis is [744, 824] arcsec and the range for y-axis is [342, 422] arcsec.}
\label{fig:maps}
\end{figure}
\end{center}

\begin{center}
\begin{figure}[pht]
\begin{tabular}{cc}
 \includegraphics[width=0.48\textwidth]{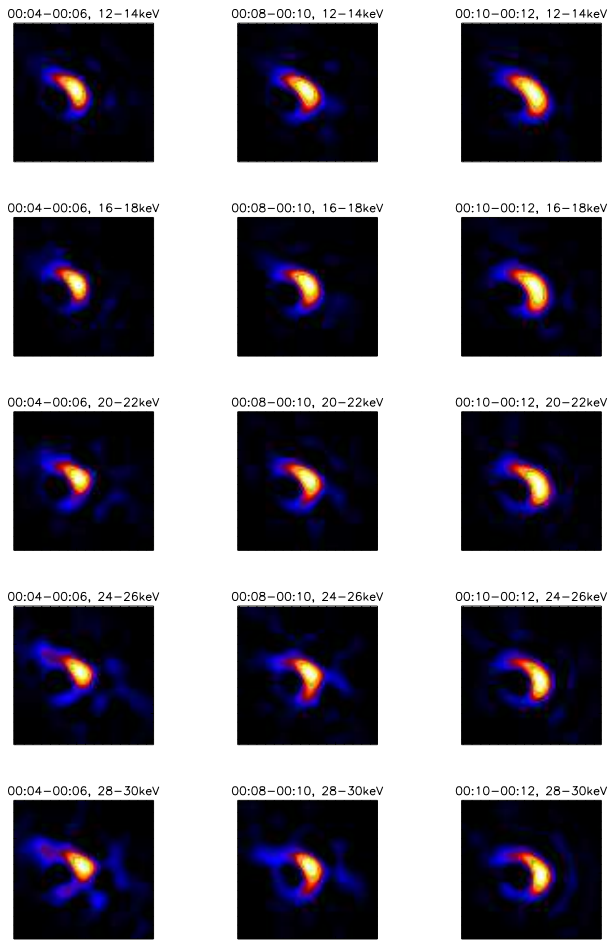}
\end{tabular}
\caption{Mean electron flux maps obtained through uv-smooth algorithm (Section \ref{uvs}) for time intervals 00:04 - 00:06 ({\it left}), 00:08 - 00:10 ({\it middle}), and 00:10 - 00:12 ({\it right})~UT. Energy bins are from 12-14~keV ({\it top}) to 28-30~keV ({\it bottom}). The ranges for x-axis and y-axis are the same as in Fig. \ref{fig:maps}.}
\label{fig:maps1}
\end{figure}
\end{center}

We have analyzed both the photon flux maps and the mean electron flux maps for five different time intervals (00:02-00:04, 00:04-00:06, 00:06-00:08, 00:08-00:10, and 00:10-00:12~UT on 2002~April~15) and ten 2-keV energy bins from 10-12~keV up to 28-30~keV. Visibilities sampled by the front detectors of collimators 3 to 9 are used for the map reconstruction \citep{xuetal08, massone2009}. Figure~\ref{fig:maps} shows the mean electron flux maps at time interval 00:06 - 00:08~UT, obtained through FWD, MEM-NJIT and UVS procedures, all using the electron visibilities obtained through spectral inversion of the observed count visibility data \citep{pianaetal07}. All three methods reveal the flare structure as a single loop. It should be noted that in general as has been shown by \cite{massone2009}, using simulated data consistent with plausible astrophysical conditions, MEM-NJIT tends to underestimate source sizes. We therefore use only FWD and UVS algorithms for the analysis to follow. Figure~\ref{fig:maps1} shows the UVS mean electron flux maps in five different energy bins (every second 2-keV energy bin from 10 to 30 keV) and three different time intervals. It can be seen that the loop length continuously grows with energy. However, with time, the size differences between low and high energies become smaller, indicating (see below) a loop density that increases with time.

\begin{center}
\begin{figure}[pht]
\begin{tabular}{cc}
\subfloat[a]{\includegraphics[width=0.45\textwidth]{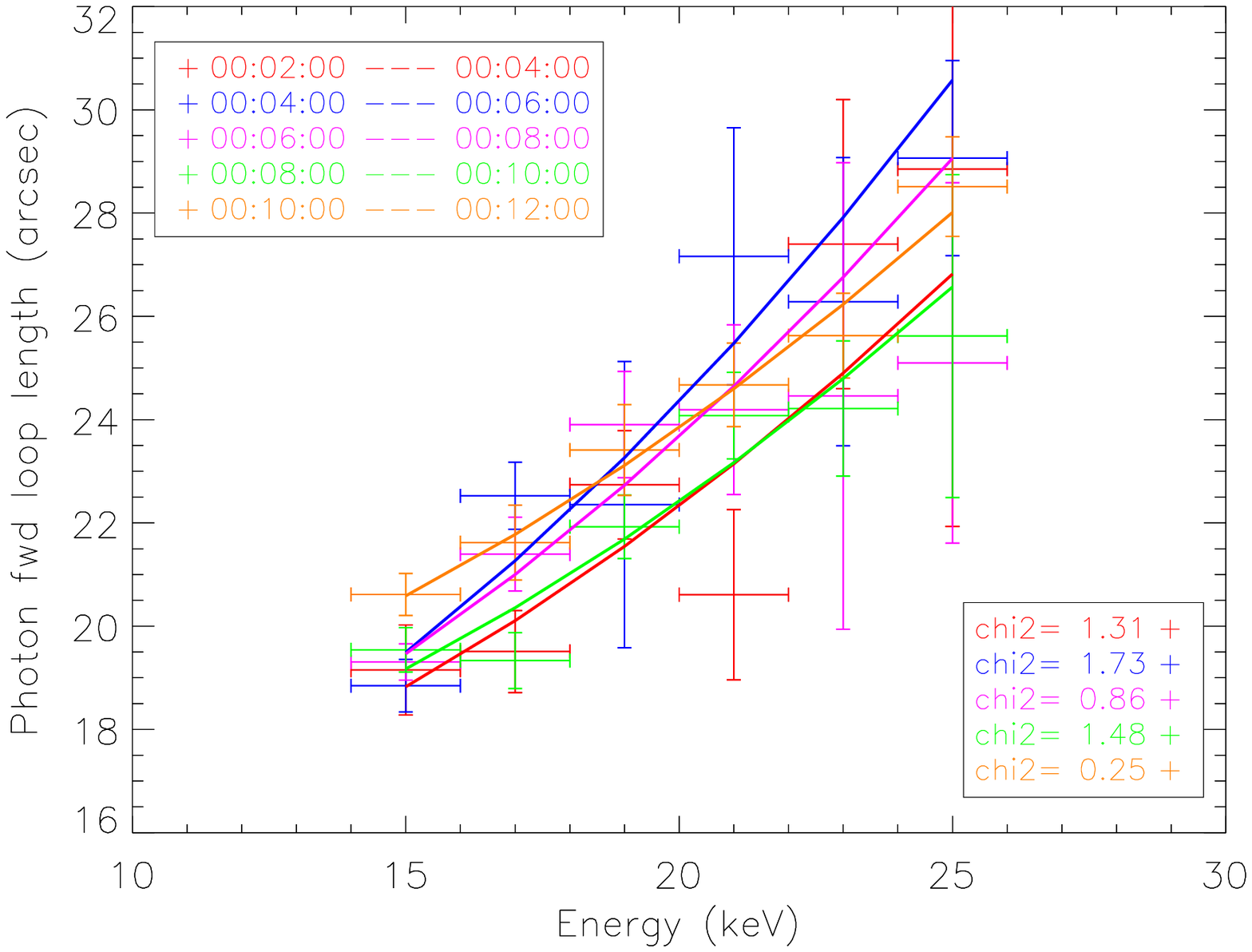}}\\
\subfloat[b]{ \includegraphics[width=0.45\textwidth]{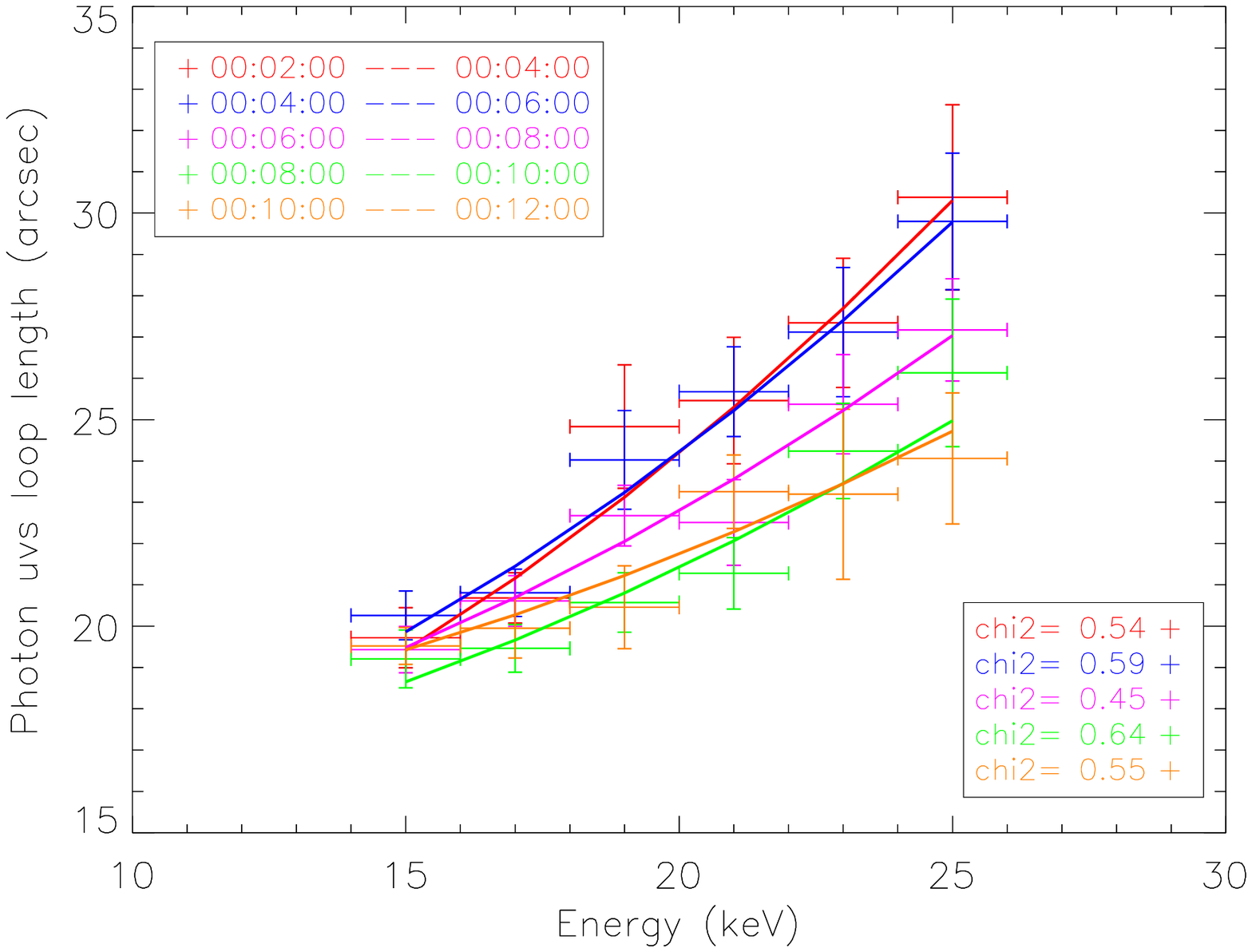}}
\end{tabular}
\caption{Flare loop lengths for each 2-keV photon energy bin, for five different time intervals throughout the flare impulsive phase. [a]: results from forward-fit algorithm; [b]: results from uv-smooth method. The solid curves represent the fits of the model described in Equation~(\ref{model-photon}). The $\chi^2$ of the fittings are also shown. }
\label{fig:photon-length}
\end{figure}
\end{center}

\begin{center}
\begin{figure}[pht]
\begin{tabular}{cc}
\subfloat[a]{\includegraphics[width=0.45\textwidth]{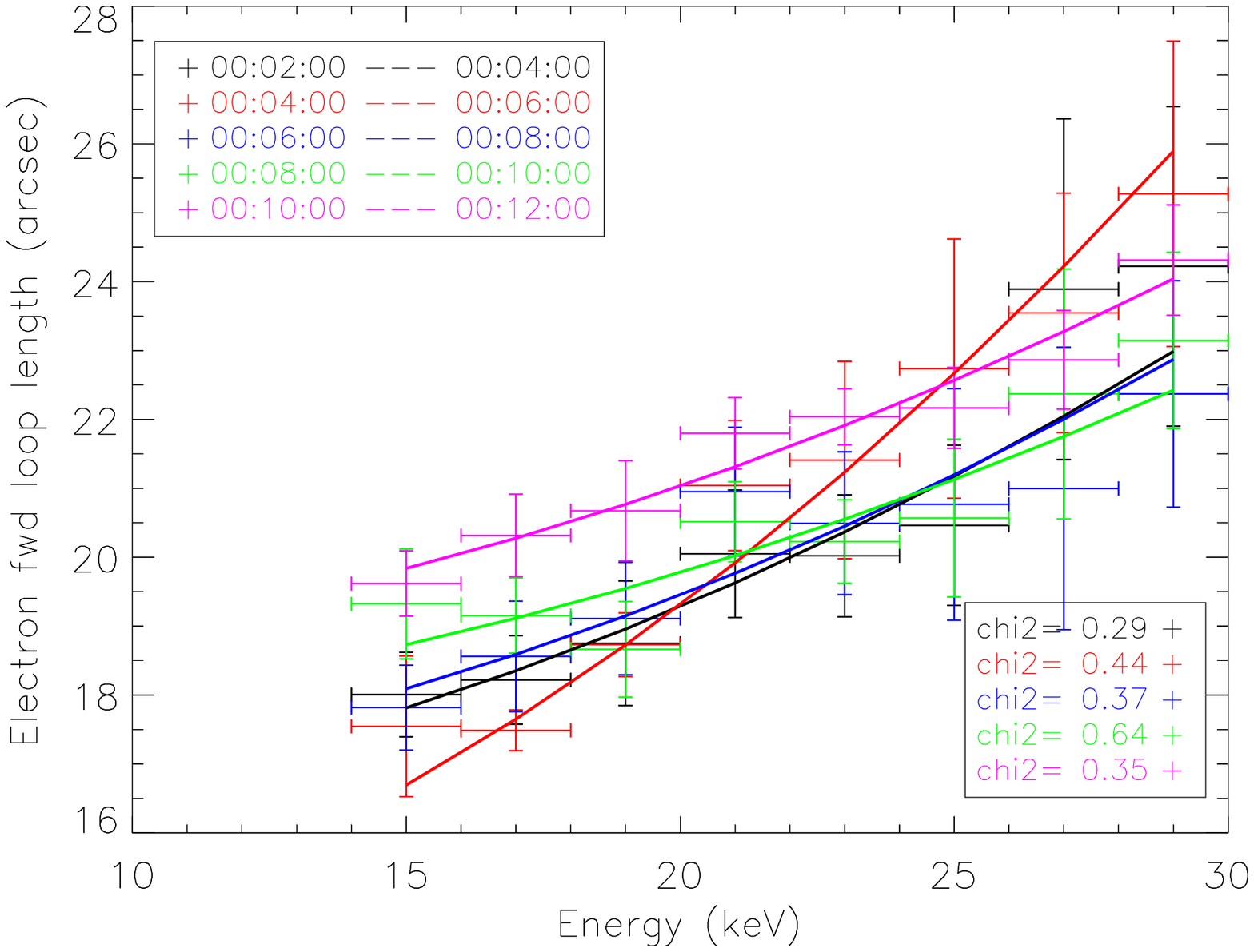}}\\
\subfloat[b]{\includegraphics[width=0.45\textwidth]{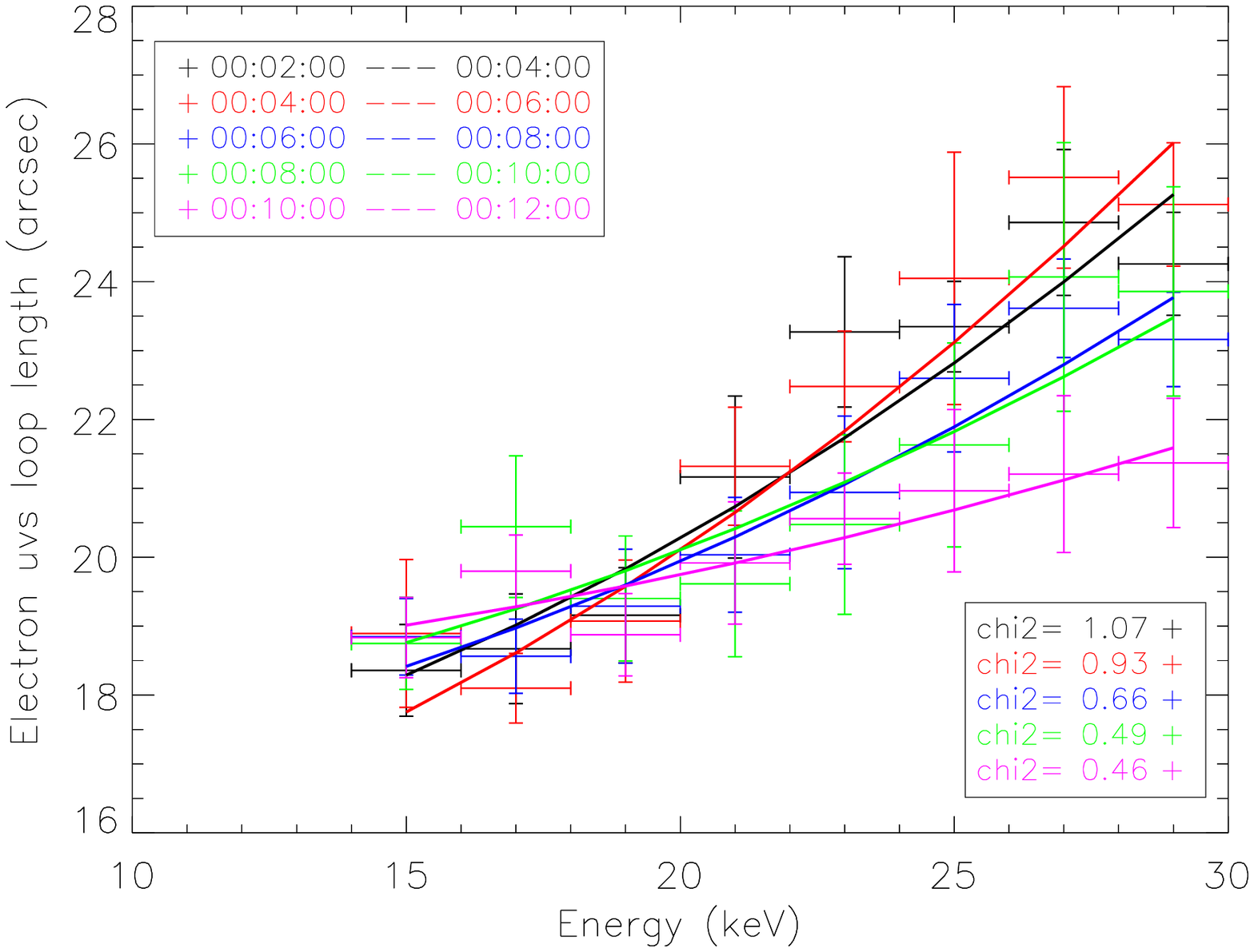}} 
\end{tabular}
\caption{Flare loop lengths for each 2-keV electron energy bin, for the same five time intervals as in Figure~\ref{fig:photon-length}. [a]: results from forward-fit algorithm; [b]: results from uv-smooth method. The solid curves represent the fits of the model described in Equation~(\ref{model-electron}). The $\chi^2$ of the fittings are also shown.}
\label{fig:electron-length}
\end{figure}
\end{center}

For the FWD imaging process we adopt a curved-elliptical-Gaussian model, for which the loop geometry is modeled by seven parameters \citep{xuetal08}. One of these seven fitting parameters, i.e., the source extent parallel to the curved arc and termed the length of the loop $L$, is particularly important to this study.  For the UVS maps, the loop length is estimated from calculating the second normalized moment of the flux intensity using Equation~(\ref{second-moment}). Figures ~\ref{fig:photon-length} and~\ref{fig:electron-length} show the lengths of the source as a function of photon energy $\epsilon$ and electron energy $E$ for five given time intervals throughout the flare impulsive phase. The loop lengths clearly grow with energy. In the ``tenuous acceleration region'' model of \citet{xuetal08}, the form of $L(\epsilon)$ is

\begin{equation}\label{model-photon}
L(\epsilon) = L_0 + \alpha  \epsilon^2,
\end{equation}
where $L(\epsilon)$ (arcsec) is the loop length at photon energy $\epsilon$ (keV); $L_0$ (arcsec) is the extent of the acceleration region, and $\alpha$ (arcsec keV$^{-2}$) is a parameter inversely proportional to the plasma density $n$ in the flare loop:

\begin{equation}\label{alphadef}
\alpha = \frac{1}{Kn} \frac{(\delta-2)}{(\delta-3)(\delta-4)}.
\end{equation}
Here $K=2 \pi e^4 \Lambda$ ($e$ being the electronic charge and $\Lambda$ being the Coulomb logarithm) and $\delta$ is the spectral index of the injected electron flux \citep{xuetal08}. This model assumes that electrons are accelerated within a tenuous region extending from $-L_0/2$ to $L_0/2$ and are injected into an external region where the loop density is sufficiently high that propogating electrons may lose energy through Coulomb
collisions and produce hard X-ray emissions \footnote{Note that a more correct ``dense acceleration region'' form for $L(\epsilon)$, which incorporates emission from the acceleration region itself, exists \citep{xuetal08}. The pertinent form of $L(E)$ is more difficult to use in a best-fit analysis; however, this model yields results for $L_0$ and $n$ that are very similar to the more straightforward-to-apply ``tenuous acceleration region'' result~(\ref{model-photon}).}. Moreover, electrons with higher energies can propagate further and hence produce photon emission over a larger spatial extent. Employing a similar analysis, we can extend the above photon-based model to the electron domain. This yields the result:
\begin{equation}\label{model-electron}
L(E) = L_0 + \beta E^{2},
\end{equation}
where $L(E)$ (arcsec) is the electron loop length at electron energy $E$~(keV) and

\begin{equation}\label{betadef}
\beta =\frac{1}{Kn} \frac{1}{(\delta - 3)} .
\end{equation}
Since the emission at photon energy $\epsilon$ is a weighted sum of electron flux at energies $E \ge \epsilon$, the overall loop extents in the photon domain (Figure~\ref{fig:photon-length}) are generally larger\footnote{From Equations [\ref{model-photon}] and [\ref{model-electron}], the ratio of the propagation lengths is ${\alpha \epsilon^2}/{\beta E^2} = ([{\delta-2}]/[{\delta-4)}]) \times  (\epsilon/E)^2$.} than those in the electron domain (Figure~\ref{fig:electron-length}).

Figure \ref{fig:ltc_map}b shows the spatially integrated \textit{RHESSI} X-ray spectrum at the third time interval (00:06 - 00:08 UT) obtained using all front detectors excluding 2 and 7 \citep{smith2002rhessi}. By fitting the spectrum with an isothermal plus collisional thick-target nonthermal model \citep{brown71}, we have found that the transition energy between the themal and nonthermal components is about 15~keV during this event. Therefore, we fit the acceleration model in both the photon and electron domains (equations [\ref{model-photon}]and [\ref{model-electron}]) starting from $E= 14$~keV. Spectral inversion of the photon-based data allows reconstruction of electron maps at energies (up to 30~keV) higher than the photon energies (up to 26~keV) observed.

The loop lengths and fits based on both FWD and UVS methods are presented in Figure~\ref{fig:photon-length} (photon domain) and Figure~\ref{fig:electron-length} (electron domain).
The best-fit acceleration region lengths $L_0$, and their statistical uncertainties (given by Monte Carlo simulations), are shown in Table~\ref{table-parameters}. For each time interval, we have also estimated the loop width $W$ averaged over different energies. Assuming $W$ is the approximate extent of the acceleration region across the magnetic loop and the loop is essentially a cylindrical column, we obtain the volume of the acceleration region to be

\begin{equation}\label{volume-acc}
V_0= \pi W^2 {L_0}/4.
\end{equation}

\begin{center}
\begin{table*}
\caption{Acceleration region length $L_0$ (Equations~[\ref{model-photon}] and~[\ref{model-electron}]), loop width~$W$ and volume~$V_0$(Equation~[\ref{volume-acc}]). }
\label{table-parameters}
\centering
\begin{tabular}{c|c|ccccc}
\hline
Time       	& (hh:mm) 				&  00:02-00:04 & 00:04-00:06 & 00:06-00:08 & 00:08-00:10 & 00:10-00:12 \\
\hline
Photon Map\\
\hline
$L_0$ (FWD)		& (arcsec) 			& $14.3\pm1.9$ &$13.3\pm1.2$ &$14.1\pm1.2$ &$15.0\pm1.0$ &$16.4\pm0.7$ \\
$L_0$ (UVS)		& (arcsec)			& $13.3\pm1.3$ &$14.3\pm1.1$ &$15.2\pm0.9$ &$15.1\pm1.1$ &$16.4\pm0.9$ \\
\hline
$W$ (FWD)		& (arcsec) 			& $7.1\pm1.1$ & $8.6\pm1.0$  & $8.1\pm0.8$ &$7.1\pm0.6$ & $6.8\pm0.5$ \\
$W$ (UVS)		& (arcsec) 			& $7.0\pm1.5$ & $7.5\pm1.3$ & $7.7\pm1.5$ & $6.6\pm1.3$ & $7.3\pm1.1$ \\
\hline
$V_0$ (FWD)	& (100~arcsec$^3$) 		& $5.7\pm1.9$ & $7.7\pm1.9$ & $7.2\pm1.6$ &$5.9\pm1.1$ & $6.0\pm0.9$ \\
$V_0$ (UVS)	& (100~arcsec$^3$) 		& $5.1\pm2.2$ & $6.3\pm2.2$ & $7.1\pm2.8$ &$5.2\pm2.1$ & $6.9\pm2.1$ \\
\hline
Electron Map\\
\hline
$L_0$ (FWD)	& (arcsec) 				& $15.9\pm0.9$ &$13.3\pm0.8$ &$16.3\pm0.9$ &$17.4\pm0.8$ &$18.3\pm0.6$ \\
$L_0$ (UVS)	& (arcsec)				& $15.7\pm0.7$ &$14.7\pm0.8$ &$16.5\pm0.6$ &$17.0\pm0.9$ &$18.1\pm0.6$ \\
\hline
$W$ (FWD)		& (arcsec) 			& $5.8\pm1.5$ & $7.8\pm1.5$  & $7.4\pm1.3$ &$6.5\pm0.8$ & $6.2\pm0.6$ \\
$W$ (UVS)		& (arcsec) 			& $7.7\pm1.8$ & $7.1\pm1.5$ & $7.4\pm1.7$ & $7.0\pm1.6$ & $7.6\pm1.6$ \\
\hline
$V_0$ (FWD)	& (100~arcsec$^3$) 		& $4.2\pm2.2$ & $6.4\pm2.4$ & $6.9\pm2.6$ &$5.7\pm1.5$ & $5.5\pm1.0$ \\
$V_0$ (UVS)	& (100~arcsec$^3$) 		& $7.3\pm3.4$ & $5.8 \pm2.5$ & $7.1 \pm3.2$ &$6.5\pm3.1$ & $8.1\pm3.4$ \\
\hline
\end{tabular}
\end{table*}
\end{center}

The values of $W$ and $V_0$, and their statistical uncertainties, are also shown in Table~\ref{table-parameters}. From these results, it can be deduced that:
\begin{itemize}
\item the models describe the data accurately in all contexts: both the photon and electron source lengths are well-fit by a quadratic form, with similar values of $L_0$ in both FWD and UVS results;
\item the acceleration region length $L_0$ is comparable to that obtained by \citet{xuetal08} ($\sim 12.5$~arcsec) using a forward-fit algorithm in the photon domain; the slight discrepancy could be due to the choice of different algorithms as well as time and energy ranges;
\item the extension of the model to the electron domain is novel, allowing smoother flux variations with energy due to the regularizing constraints involved in the inversion process; this in turn validates the photon-based results and represents one step closer to the underlying physics. The electron maps generally offer a better fit (i.e., the $\chi^2$ values are smaller), particularly using the FWD procedure.
\item the loop lengths appear to show a systematic change throughout the event: the energy-length curves are steeper at earlier time intervals and flatter at later time intervals, especially in the electron domain shown in Figure~\ref{fig:electron-length}. According to the model (Equation [\ref{betadef}]) this may be due to the increasing density as the flare evolves, given that the electron spectral index $\delta$ does not show a systematic change with time \footnote{The spectral index of the injected electron flux  $\delta$  is obtained from \textit{RHESSI} X-ray spectra fitting using an isothermal plus thick-target nonthermal model. Through the five studied time intervals, $\delta$ has values as [7.6, 7.3, 7.9, 8.6, 7.8] }.
\end{itemize}

\section{Interpretation of the Results}\label{result}

It is often postulated that electrons are accelerated at a reconnection current sheet and that the acceleration process is decoupled from the electron transport that follows the reconnection \citep{asch1998deconvolution}. Indeed, many solar flares demonstrate a clear loop-top-plus-footpoints structure \citep[see, e.g.,][]{Battaglia2011}. However, this flare has a loop density as high as $n \simeq 10^{11}$~cm$^{-3}$ \citep{vebr04}. Therefore, most of the energetic electrons would be stopped through Coulomb collisions near the looptop \citep{kohabi11}, and so the emission region would be confined to a limited extent less than the loop length ($\sim30\arcsec$). The spatially resolved observations presented here suggest the presence of an extended acceleration site inside the loop.  Further, given that the acceleration region ($\sim15\arcsec$) is estimated to be a substantial fraction of the total length of the flare loop ($\sim30\arcsec$; see Figures \ref{fig:photon-length} and  \ref{fig:electron-length}) and that the hard X-ray emission is produced in the entire loop instead of only at the footpoints, it is clear that the dense flare loop incorporates both acceleration and transport of electrons, with concurrent thick-target bremsstrahlung emission over the entire length of the source.

Such a scenario is consistent with the presence of an enhanced level of MHD fluctuations inside the loop \citep{bian2011}. It is possible that for such events reconnection and loop-top injection may not correspond to the dominant energy dissipation and particle acceleration process, which instead may proceed inside the flare loop itself through stochastic mechanisms \citep[e.g.,][]{petrosian_liu2004}.  Our result shows that a considerable proportion of the flare loop is involved in the process of acceleration, thus avoiding the ``number problem'' \citep{brown2009}.

\section{Conclusions}\label{conclusion}

We have analyzed an extended coronal hard X-ray source in the event of 2002~April~14 with photon maps constructed from count visibilities (two-dimensional spatial Fourier transforms of the source geometry).
Furthermore, using a regularized spectral inversion technique generating electron visibilities, we have studied the dependence of the electron flux image on electron energy $E$. The source lengths derived from both photon and electron maps generally grow quadratically with energy, in agreement with a collisional model involving an extended acceleration region \citep{xuetal08, kohabi11,bian2011}.
Fitting this model allows estimation of the length and volume of the acceleration region: $\sim 15$ arcsec and $\sim600$ arcsec$^3$, respectively.
We compare the results obtained by different algorithms (FWD and UVS) and in both photon and electron visibility domains. The plausible and consistent estimates of the acceleration-region lengths and widths strongly suggest that the proposed model is reliable.
The systematic change of the behavior of loop length with time indicates that the loop density may increase throughout the event, due to chromospheric "evaporation" \citep{acton1982}. The inclusion of a thermal component to the current model is required for a comprehensive description of the chromospheric evaporation process.

Detailed and statistical studies of more cases of dense-loop flares will be carried out in future work, in which we will also investigate other properties of the acceleration region, e.g., the loop density and its evolution with time.

\begin{acknowledgements}
The authors are very grateful to the referee for helpful comments and we would like to thank  Richard Schwartz, Kim Tolbert, Gordon Hurford, Iain Hannah and Yan Xu for useful \textit{RHESSI} techniques and fruitful discussions. The work is supported by STFC Rolling grant and the EU FP7 HESPE grant no. 263086; AGE was supported by NASA Grant NNX10AT78J.
\end{acknowledgements}

\bibliographystyle{aa}
\bibliography{guo_bib}

\end{spacing}
\end{document}